\documentclass[preprint,showpacs,preprintnumbers,amsmath,amssymb]{revtex4}

\usepackage{graphicx}
\usepackage{dcolumn}
\usepackage{bm}
\newcommand{\be}{\begin{equation}}
\newcommand{\ee}{\end{equation}}
\begin{document}


\title{Evidence of thin-film precursors formation in hydrokinetic and atomistic simulations of nano-channel capillary filling} 

\author{S. Chibbaro}\affiliation{Istituto Applicazioni Calcolo, CNR, V.le del Policlinico 137, 00161, Rome, Italy, e-mail:chibbaro@iac.cnr.it}
\author{L. Biferale}\affiliation{University of Tor Vergata and INFN, via della Ricerca Scientifica 1, 00133 Rome, Italy}
\author{F. Diotallevi}\affiliation{Istituto Applicazioni Calcolo, CNR, V.le del Policlinico 137, 00161, Rome, Italy}
\author{S. Succi}\affiliation{Istituto Applicazioni Calcolo, CNR, V.le del Policlinico 137, 00161, Rome, Italy}
\author{K. Binder}\affiliation{Institut fur Physik, Johannes Gutenberg Universitat Mainz, Staudinger Weg 7, 55099 Mainz Germany}
\author{D. Dimitrov}\affiliation{Institute for chemical physics, Bulgarian Academy of Sciences, 1113 Sofia, Bulgaria}
\author{A. Milchev}\affiliation{Institut fur Physik, Johannes Gutenberg Universitat Mainz, Staudinger Weg 7, 55099 Mainz Germany}
\author{S. Girardo}\affiliation{CNR c/o Universita' degli Studi di Lecce via Arnesano - I-73100 Lecce (Italy)}
\author{D. Pisignano}\affiliation{CNR c/o Universita' degli Studi di Lecce via Arnesano - I-73100 Lecce (Italy)}

\begin{abstract}
We present hydrokinetic Lattice Boltzmann and Molecular Dynamics
simulations of capillary filling of high-wetting
fluids in nano-channels, which provide clear evidence of the formation
of thin precursor films, moving ahead of the main capillary front.
The dynamics of the precursor films is found to obey the Lucas-Washburn 
law as the main capillary front, $z^2(t) \propto t$, although with a larger prefactor,
which we find to take the same value for both geometries under inspection.
Both hydrokinetic and Molecular Dynamics approaches indicate
a precursor film thickness of the order of one tenth of the capillary diameter.
The quantitative agreement between the hydrokinetic and atomistic methods
indicates that the formation and propagation of thin precursors can be handled
at a mesoscopic/hydrokinetic level, thereby opening the possibility of using
hydrokinetic methods to
space-time scales and complex geometries of direct experimental relevance.
\end{abstract}

\maketitle

Micro-  and nanohydrodynamic flows  are playing  an emerging  role for
many  applications in  material  science, chemistry  and biology,  and
consequently  they cast  a pressing  demand  for a  deeper and  better
understanding  of the  basic mechanisms  of  fluid flow  at micro  and
nanoscales~\cite{DeG_84,DeG_book,Bhu_95,Str_02,Sup_03,Yak_07,Jos_06},
as   well  as   for   the  development   of  corresponding   efficient
computational  tools.   The  formation  of  thin  precursor  films  in
capillary  experiments with highly  wetting fluids  (near-zero contact
angle) has  been reported by  a number of experiments  and theoretical
works~\cite{Bonn,Kav_03,Bico},  mostly   in  connection  with  droplet
spreading  experiments.   In spite  of  its  conceptual and  practical
importance, however, much less work  has been directed to the study of
precursor films for the case of capillary filling.  Indeed, this is of
great interest because,  by forming and propagating ahead  of the main
capillary  front,   thin-film  precursors  may  manage   to  hide  the
chemical/geometrical  details   of  the  nanochannel   walls,  thereby
exerting a  major influence  on the efficiency  of nanochannel-coating
strategies \cite{Cottin,tas}.  
In this realm, the continuum assumption  behind the macroscopic description of
fluid flow  goes often  under question, typical  cases in  point being
slip-flow at solid walls and moving contact line of liquid/gas interface
on solid walls~\cite{DeG_84,Bonn}.
In  order  to   keep  a  continuum
description  at nanoscopic  scales and  close to  the  boundaries, the
hydrodynamic equations are  usually enriched with generalised boundary
conditions, designed in  such a way as to  collect the complex physics
of fluid-wall  interactions into a  few effective parameters.
A more radical
approach  is to  quit the  continuum level  and turn  directly  to the
atomistic  description  of  fluid  flows  as a  collection  of  moving
molecules~\cite{Frenkel}.
This  approach is computationally demanding,
thus preventing the attainment of space and time macroscopic scales
of  experimental interest.   In recent  years, a  third, intermediate,
alternative is  rapidly materialising, in the form  of minimal lattice
versions  of the Boltzmann  kinetic equation  \cite{SS_92,wolf}.  
By definition, such  a mesoscopic approach  is best suited  to situations
where  molecular  details,  while  sufficiently important  to  require
substantial amendments  of the  continuum assumption, still  possess a
sufficient   degree  of  universality   to  allow   general  continuum
symmetries   to  survive,  a   situation  that   we  shall   dub  {\it
supra-molecular}  for   simplicity.   Lacking  a   rigorous  bottom-up
derivation,   the   validity  of   the   hydro-kinetic  approach   for
supra-molecular physics must be assessed case-by-case, a program which
is    already    counting     a    number    of    recent    successes
\cite{SS_06,Sbr_06,Yeom,Harting}.  
The  aim of this paper  is to study
another potentially 'supramolecular' situation, i.e. the formation and
propagation  of precursor  films  in capillary  filling at  nanoscopic
scales. 
We  provide  quantitative evidences  that  the
formation  and  propagation  of   a  precursor  film  is  governed  by
kinetic-hydrodynamical laws, and that the emergent asymptotic dynamics
for the head of the film follows  a square root law --in time-- as the
bulk  fluid. Nevertheless, the  prefactors of  the two  behaviours are
different,  eventually  leading  to  a  stable  infinitely  long  thin
precursor film ahead of the front.  We employ both MD and hydrokinetic
simulations, finding  quantitative agreement for  both bulk quantities
and local density profiles, at  all times during the capillary filling
process.   This  lends  further   support  to   the  idea   that  many
supramolecular   problems   are     efficiently   handled   with
mesoscopic/hydrokinetic methods. 

\begin{figure}
\vspace{0.5cm}
(a)
\includegraphics[scale=0.35]{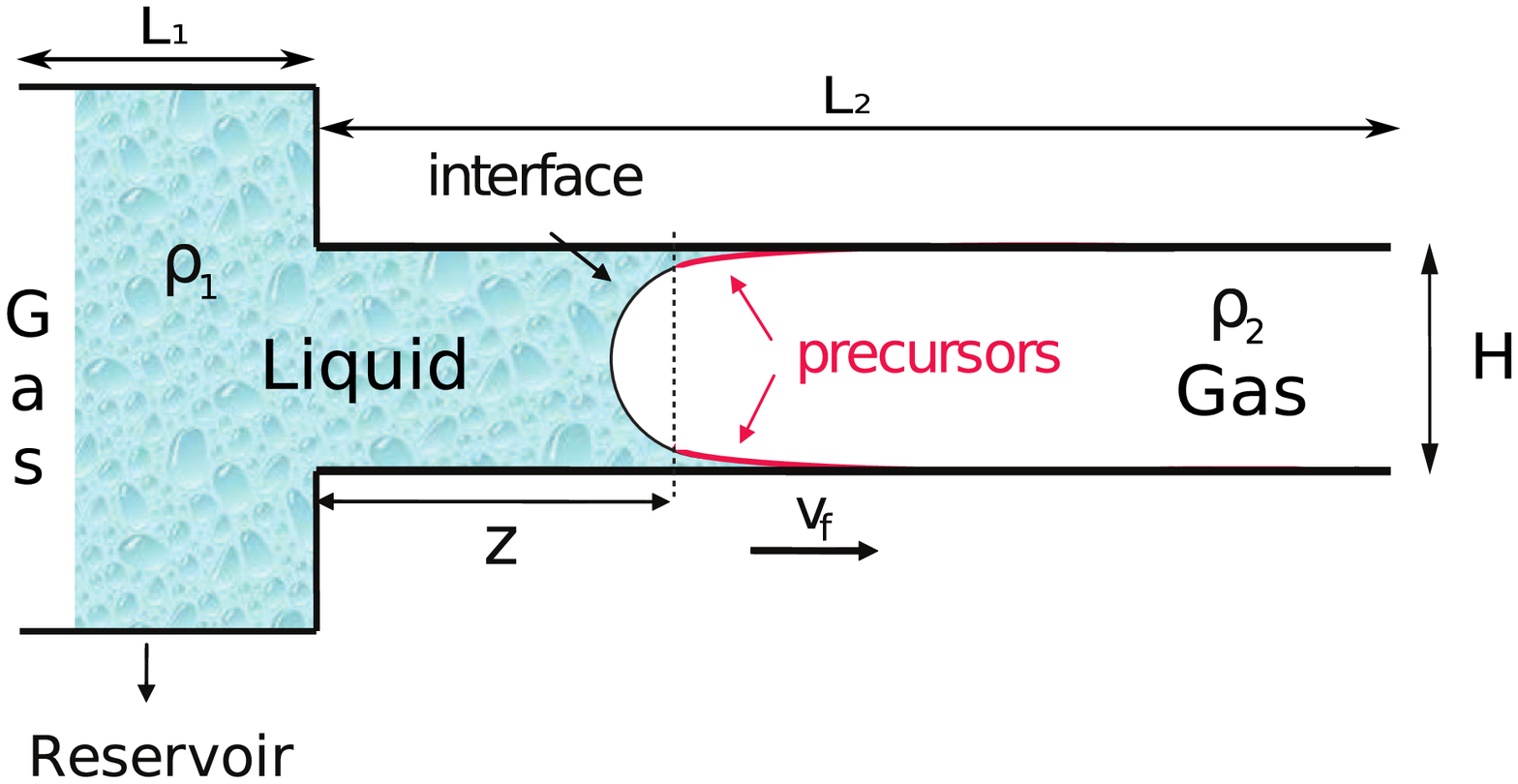}\hspace{0.25cm}
(b)
\includegraphics[scale=0.45]{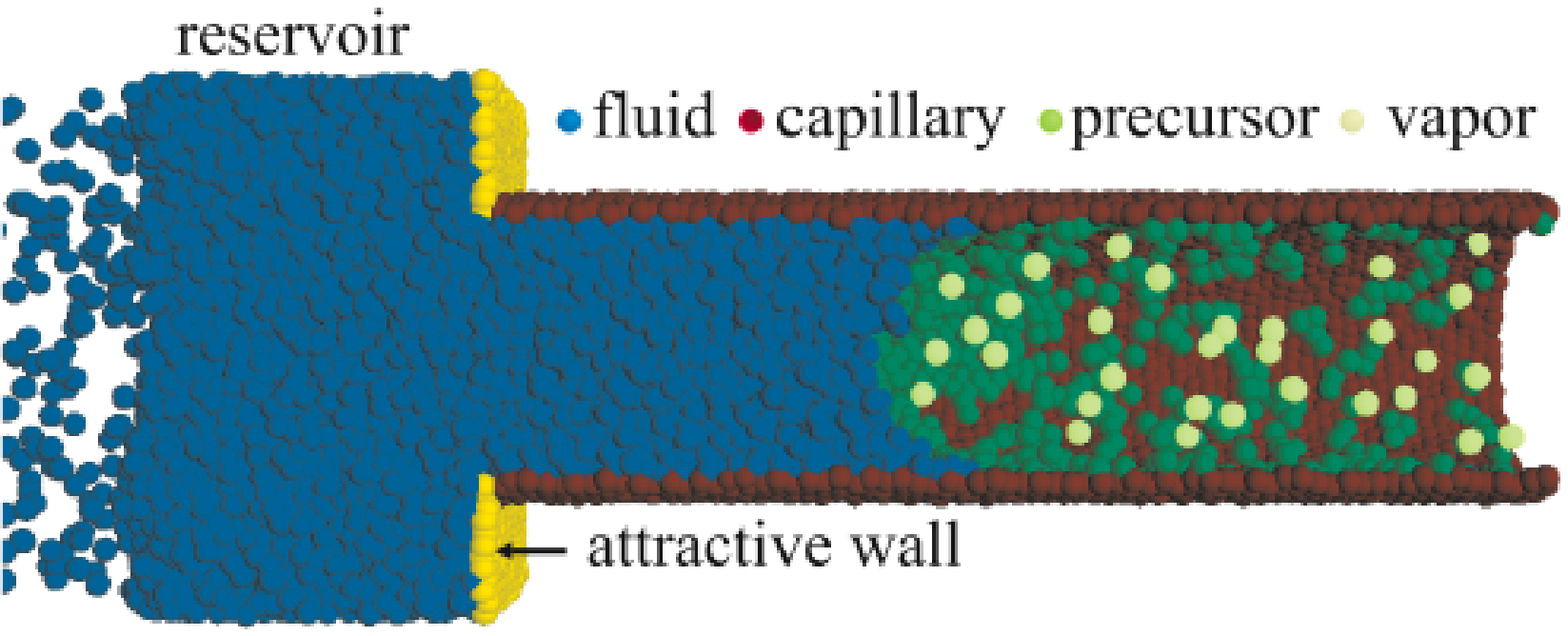}\hspace{0.25cm}
\caption{Sketch of the geometry used for the description of the capillary imbibition in the LB and MD simulations. (a) The 2 dimensional geometry, with length $L_1+L_2$ and width $H$, is divided in two parts. The left part has top and bottom periodic boundary conditions, so as to support a perfectly flat gas-liquid interface, mimicking an ``infinite reservoir''. In the right half, of length $L_2$, there is the actual capillary: the top and bottom boundary conditions are solid wall, with a given contact angle $\theta$. Periodic boundary conditions are also imposed at the inlet and outlet sides. The main LB parameters are: $H\equiv ny=40, L_1=nz=170, \rho_1=1; \rho_2=0.35, \mu_1=0.66, \mu_2=0.014, \gamma=0.016$ where $H$ is the channel height, $L_2$ is the channel length, $\rho_2$ and $\rho_1$ the gas and liquid densities respectively: $\mu_k$, $k=1,2$ the dynamic viscosities, and $\gamma$ the surface tension.  (b) Snapshot of fluid imbibition for MD in the capillary at time $t=1300$ MD time-steps. The fluid is in equilibrium with its vapour. Fluid atoms are in blue. Vapour is yellow, tube walls are red and the precursor is green. One distinguishes  between vapour and precursor, subject to the radial distance of the respective atoms from the tube wall, if a certain particle has no contact with the wall, it is deemed 'vapour'.  The MD parameters are as follows \cite{Dim_07}: $R=11 \sigma, L=55 \sigma, \rho_l=0.774, ~\mu=6.3,~\gamma=0.735,~\sigma=1$, where $R$ is the capillary radius and $L$ its lenght.}
\label{Fig:1}
\end{figure}

In this work, we use the multicomponent LB model proposed by Shan and Chen \cite{SC_93}. 
This model caters for an arbitrary number of components, with different molecular mass and reads
as follows:
\begin{equation}
\label{eq:lbe}
f^k_i(\vec x+ {\bm c}_{i} \Delta t,t+\Delta t)-f^k_i(\vec x,t)=-\frac{\Delta t}{\tau_k} 
[f^k_i(\vec x,t) -f_i^{k(eq)}(\vec x,t)] 
\end{equation}
where $f^k_i(\bm{x},t)$ is the  probability density function
associated with a mesoscopic particle of velocity $\bm{c}_{i}$ for the $k$th component,
 $\tau_k$ is a mean
collision time of the $k$th component
($\Delta t$ being the time-step associated with particle free-streaming), and $f^{k(eq)}_{i}(\vec x,t)$
is the corresponding Maxwell equilibrium function.
The interaction force between particles is the sum of
a bulk and a wall components:
\be
\label{forcing} 
{\bf F}_{bk}({\vec x})=-n_k(\vec x)\sum_{\vec x} \sum_{\bar{k}=1}^sG_{k \bar{k}}n_{\bar{k}} (\vec x)(\vec x+\Delta \vec x);\;\;
{\bf F}_{wk}({\vec x})=-n_k(\vec x)\sum_{\vec x} g_{k w} n_w
 (\vec x)(\vec x+\Delta \vec x)
\ee
where $n_k$ is the local density and the interaction matrix $G_{k\bar{k}}$ is symmetric  .
In our model,  $G_{k \bar{k}}= g_{k \bar{k}}, (\textrm{for}~ |\vec x^{\prime}-\vec x|=1);$,
 $G_{k \bar{k}}=g_{k \bar{k}}/4, (\textrm{for}~|\vec x^{\prime}-\vec x|=\sqrt{2};$ and 
$0~ \textrm{otherwise}$.
At the fluid/solid interface, the wall is regarded as a (solid) phase with a constant number density
 $n_w$  and $g_{kw}$ is the interaction 
strength between component $k$ and the wall. 
By adjusting the parameters $g_{kw}$ and $n_w$ , different wettability can be obtained~\cite{Kan_02}. 
This approach allows the definition of a static contact angle $\theta$,
spanning the entire range $\theta \in [0^o:180^o]$~\cite{ben_06}.
To the purpose of analysing the physics of film precursors, it is important to 
notice that , in the limit where the hard-core repulsion is negligible,
both the Shan-Chen pseudo-potential and van der Waals interactions
predict a  non-ideal equation of state, in which the leading  correction to the ideal pressure is $\propto \rho^2$.
Concerning the MD simulation~\cite{Dim_07}, see fig. \ref{Fig:1}, 
fluid atoms interact through a LJ potential,
$U_{LJ} (r) = 4\epsilon[(\sigma/r)^{12} - (\sigma/r)^6]$, where $\epsilon =
1.4$ and $\sigma = 1.0$. The capillary walls are represented by atoms forming a triangular
 lattice with spacing 1.0 in units of the
liquid atom diameter $\sigma$. The wall atoms may fluctuate
 around their equilibrium positions at $R + \sigma$, subject
 to a finitely extensible non-linear elastic (FENE) potential $U_{FENE} = −15\epsilon_wR_0^2ln(1-r^2/R_0^2)$, $R_0=1.5$.
Here $\epsilon_w = 1.0k_BT,$ where  $k_B$ denotes the Boltzmann constant and T is the temperature of the system. In
addition, the wall atoms interact through a LJ potential,
$U_{LJ} (r) = 4\epsilon_{ww}[(\sigma_{ww}/r)^{12} − ((\sigma_{ww}/r)^6]$, where $\epsilon_{ww} =
1.0$ and $\sigma_{ww} = 0.8$.
This choice  guarantees
no penetration of liquid particles through the wall. 
Molecules are advanced in time via the velocity-Verlet algorithm and a
dissipative-particle-dynamics thermostat, with friction parameter $\xi = 0.5$,
and a thermostat cutoff $r_c=2.5\sigma$~\cite{Frenkel}.
The integration time step is $\delta t = 0.01t_0,$ where
 $t_0 = \sqrt{m\sigma^2/48k_BT}=1/\sqrt{48}$
is the basic time unit and we have taken
the particle mass $m \equiv 1$ and $k_BT \equiv 1$.

We consider a capillary filling experiment, whereby a dense fluid, of density and dynamic viscosity $\rho_1,\mu_1$, 
penetrates into a channel filled up by a lighter fluid, $\rho_2,\mu_2$, see fig. \ref{Fig:1}.
For this kind of fluid flow, the Lucas-Washburn law is expected to hold, at least at macroscopic scales~\cite{Wash_21,Lucas} and in the limit $\mu_1\gg \mu_2$.
Recently, the same law has been observed 
even in nanoscopic experiments \cite{Hub_07}. 
In these limits, the law governing the position, $z(t)$, of the macroscopic meniscus is: 
$
z^2(t) - z^2(0) = \frac{\gamma H cos(\theta)}{C \mu }   t ,$
 where $\gamma$ is the surface tension between liquid and gas, $\theta$
is the {\it static} contact angle, $\mu$ is the liquid viscosity, $H$
is the channel height and the factor $C$ depends on the flow geometry (in the present geometry $C_{LB}=3\,;\,C_{MD}=2$).
The geometry we are going to investigate is depicted in
fig. (\ref{Fig:1}) for both models.  It is important to underline that
in the LB case, we simulate two immiscible fluids, without any phase transition. 


Since binary LB methods do not easily support high density ratios between
the two species, we impose the correct ratio between
the two dynamics viscosities through an appropriate choice of the
kinematic viscosities. The chosen
parameters correspond to an averaged capillary number $Ca \approx 3 \;
10^{-2}$ and $Ca \sim 0.1$ for LB and MD respectively. 

\begin{figure}
\vspace{0.5cm}
(a)
\includegraphics[height=5cm,width=7.cm]{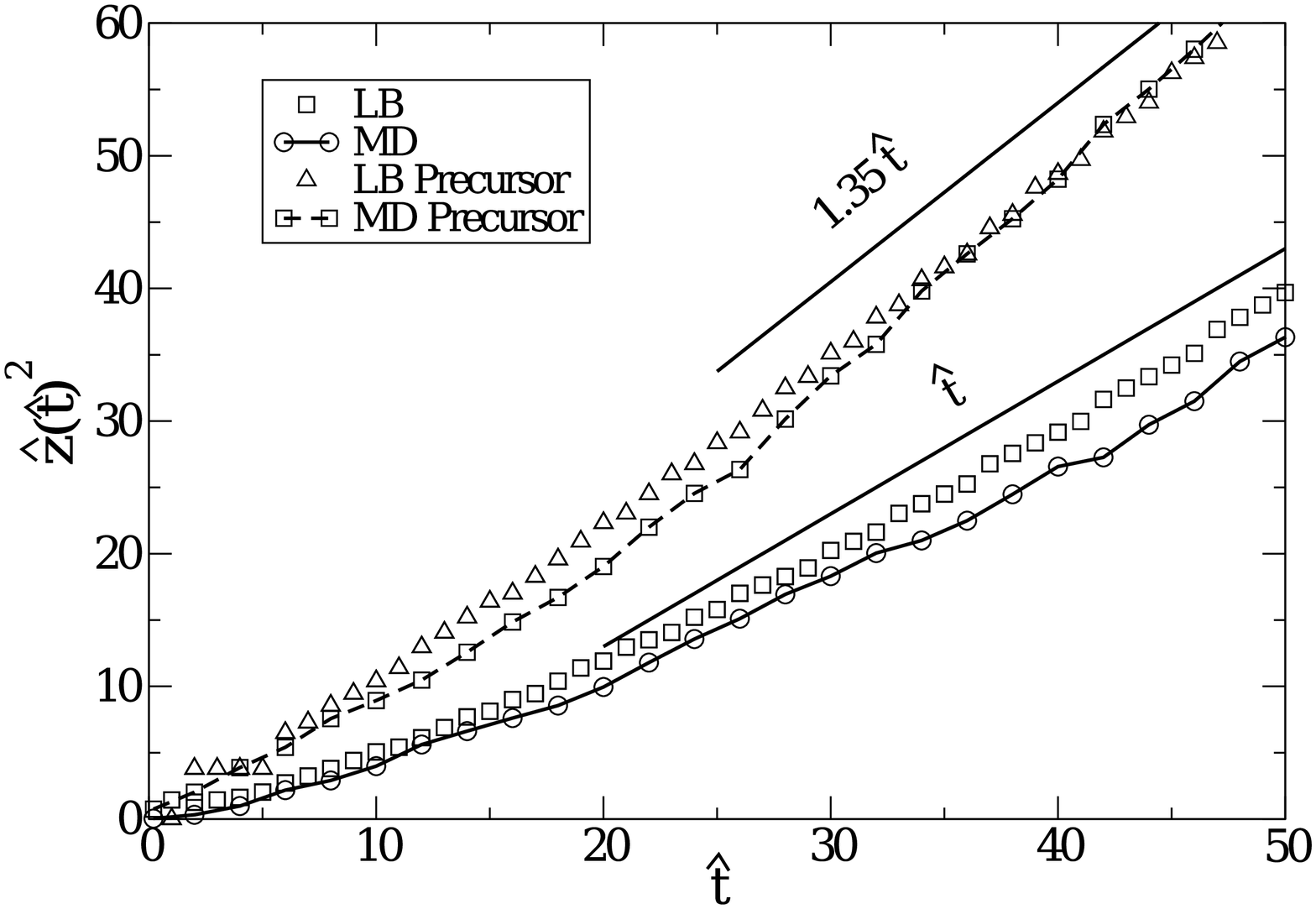}\vspace{1cm}\\
(b)
\includegraphics[height=5cm,width=7.cm]{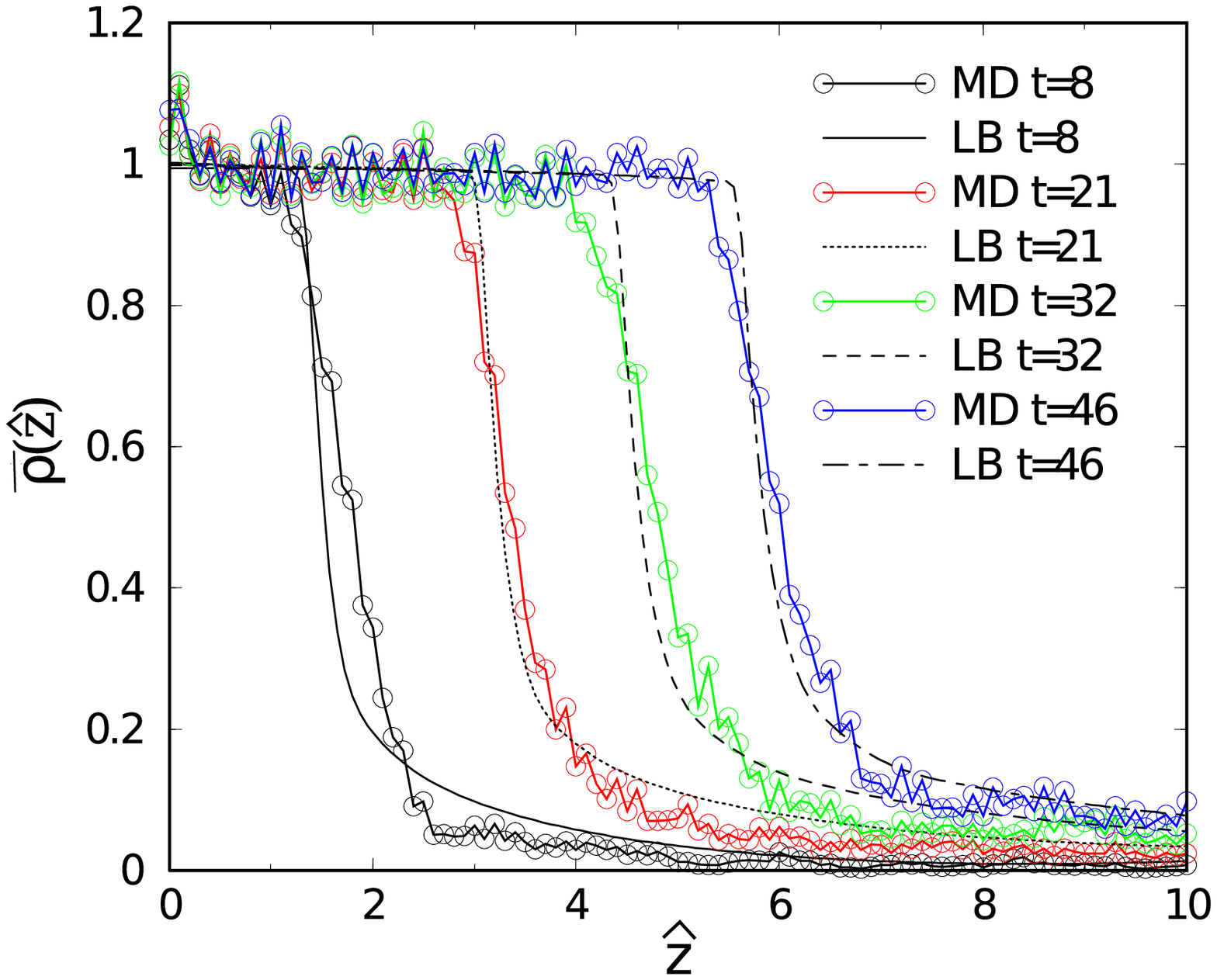}
\caption{ Dynamics of the bulk and precursor meniscus. (a) Position of the liquid meniscus $\hat{z}^2(t)$  for LB and MD simulations. The position
  of the precursor film, $\hat{z}^2_{prec}(t)$ is also plotted for both
  models.  $\hat{z}_{prec}$ is defined as the rightmost location with density
  $\rho=\rho_{bulk}/3$.  All quantities are given in natural ``capillary''
  units (see text). The asymptotic ($t > 15 t_{cap}$) rise of both precursor and
  bulk menisci follows a Lucas-Washburn law, with different prefactors
  (see the two straight lines). Notably, the precursor film is found to proceed with the law:$\hat{z}^2_{prec}(t)=1.35 \hat{t}$.  (b) Profiles of the average fluid
  density $\overline{\rho}(z)$ in the capillary at various times for
  LB and MD models.}
\label{Fig:2}
\end{figure}

In order to expose the universal character of the phenomenon, results
are presented in natural units, namely, space is measured in units of
the capillary size, $l_{cap}$ and time in units of the capillary time
$t_{cap}=l_{cap}/V_{cap}$, where $V_{cap}=\gamma/\mu$ is the capillary
speed and $l_{cap}=H/C_{LB}$ for LB and $l_{cap}=R/C_{MD}$ for MD. In
these units, the Lucas-Washburn law takes a very simple universal form 
\begin{equation}
\label{eq:univ}
{\hat   z}^2=\hat{z}_0^2+ \hat{t}~, 
\end{equation}
 where we have inserted the value of 
$cos(\theta)=1$, corresponding to complete wetting.
As to the bulk front position, fig.\ref{Fig:2}a shows that both MD and LB
results superpose with  
the law (\ref{eq:univ}), while the precursor  position develops a faster dynamics, fitted
by the relation: 
\begin{equation}
{\hat   z}^2_{prec}=\hat{z}_0^2+ 1.35 \hat{t}~.
\label{eq:prec}
\end{equation}
 Similar speed-up of the precursor has  been reported also in different experimental situations \cite{Bico}. The precursor is here defined through the density profile, $\overline{\rho}(z)$,  averaged
over the direction across the channel. 
In figure \ref{Fig:2}b, we show $\overline{\rho}(z)$
at various instants (always in capillary units), for both MD and LB
simulations.  
The relatively high average density $\overline{\rho}(z)$
near  the wall, at later times, witnesses the presence of a
precursor film attached to the wall.  From this figure, it is
appreciated that quantitative agreement between MD and LB is found also at
the level of the spatial profile of the density field.  This is
plausible, since the LB simulations operate on similar principles as
the MD ones, namely the fluid-wall interactions are {\it not}
expressed in terms of boundary conditions on the contact angle, like
in continuum methods, but rather in terms of fluid-solid
(pseudo)-potentials.  In particular, the degree of hydrophob/philicity
of the two species can be tuned independently, by choosing different
values of the fluid-solid interaction strengths (for details see
\cite{Dim_07}).  In figure \ref{Fig:2}b, we show the position of
the advancing front and  the precursor film as a function of time,
for both MD and LB simulations.  Even though the averaged capillary
numbers are not exactly the same, a pretty good agreement between LB and MD is observed.  In
particular, in both cases, the precursor is also found to obey a $\sqrt
t$ scaling in time, although with a larger prefactor than the front.
 As a result, the relative distance between the two
keeps growing in time, with the precursors serving as a sort of 'carpet',
hiding the chemical structure of the wall to the advancing front.  
As recently pointed out~\cite{Dim_07,Hub_07}, MD
simulations and experiments indicate that the Lucas-Washburn law persists
down to nanoscopic scales. The fact that the precursor dynamics 
  {\it quantitatively} matches with mesoscopic simulations, suggests
  that the precursor physics also shows the same kind of nanoscopic
  persistence. 
\begin{figure}
\vspace{0.5cm}
(a)
\includegraphics[scale=0.3]{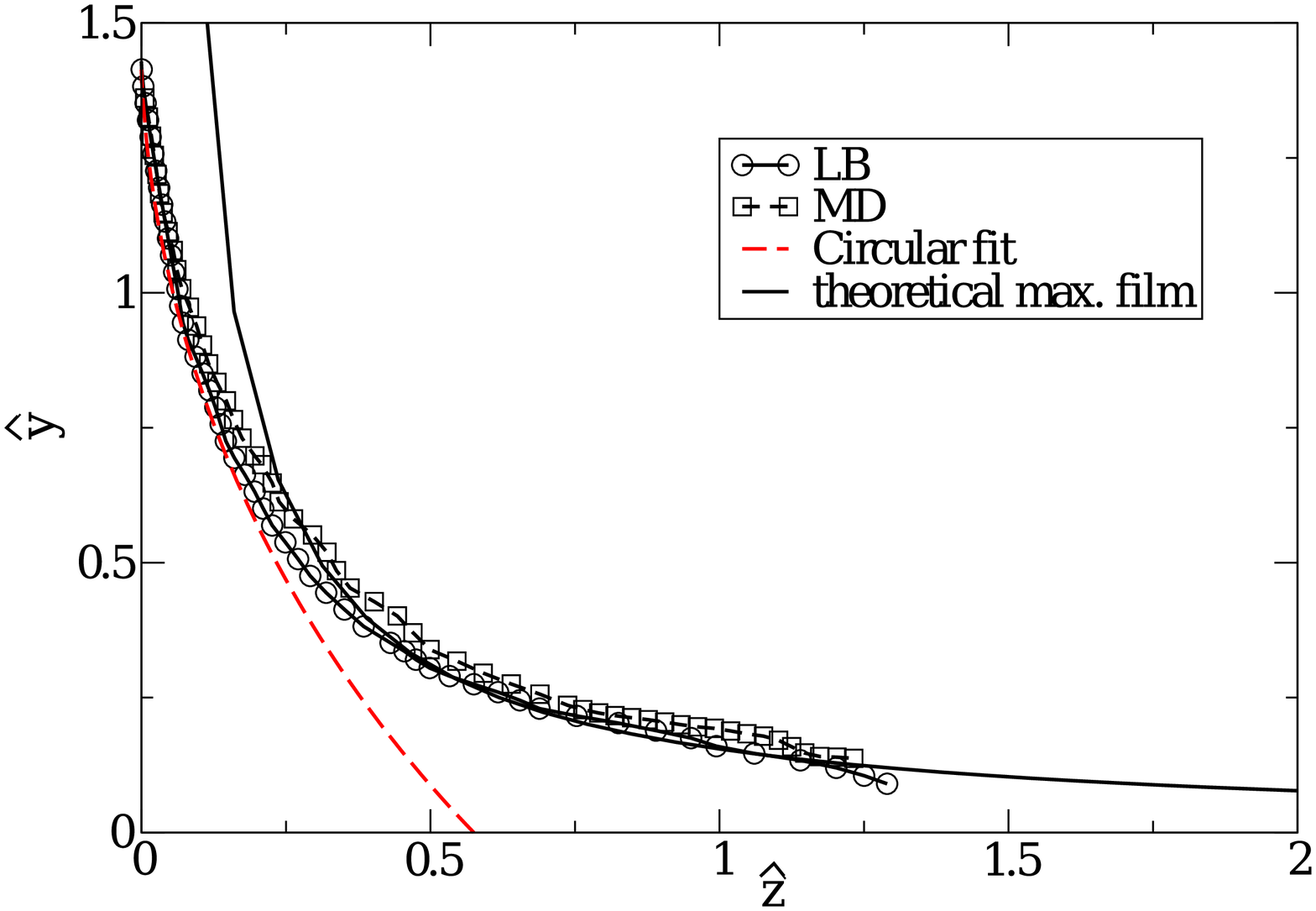}
(b)
\includegraphics[scale=0.275]{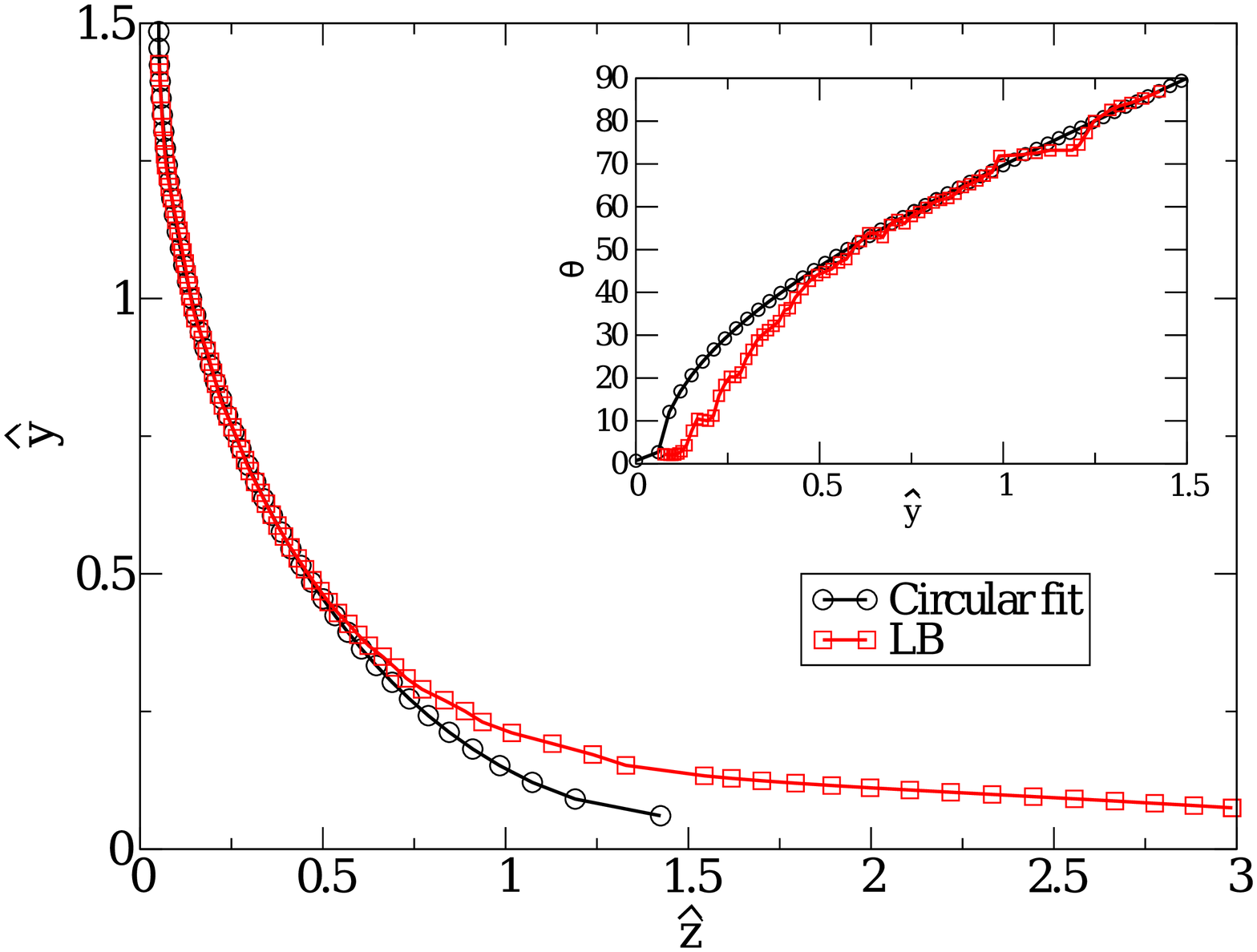}
\caption{Structure of the precursor film. (a) Snapshot of the interface profile for LB and MD models at
  time $\hat{t}=20$. Both LB and MD profiles are obtained as isolines
  at $\rho_{bulk}/\rho=3$. We also show the circular fit (bold
  line), valid for the menisci in the bulk region and the $1/z$
  profile as suggested by lubrication theory. Notice the clear
  departure of the interface form a circular fit, due to the
  different physics close to the wall.  Natural ``capillary'' units are used (see text).
  (b): The interface profile at
  a later stage of capillary filling (symbol: square). The circle-straight line  corresponds
  to the circular fit of the bulk ($0.15<\hat{y}<1.5$). Notice the long
  extension of the precursor, here defined as the spatial interval where the
  profile departs from the circular fit.  The length of the precursor
  film is consistent with the relation $L_p \sim Ca^{-1}$, as found in
  a recent nanoscale experiment~\cite{Kav_03} (not shown). In the inset, we plot
the value of the macroscopic angle defined by the interface curvature at varying the distance from the wall.}
\label{Fig:4}
\end{figure}
Concerning the instantaneous interface profile,  in Fig. \ref{Fig:4}a
we show the structure of the precursor by analysing 
fluid menisci cuts for different density levels. The
agreement between the two methods is again quantitative.  
In LB, the interface turns out to be about $5$ $\Delta z$
(leading to the rough estimate, $\Delta z \approx 0.2 nm$, considering an
interface thickness for water/air at room temperature  $\Delta \xi \approx 1nm$). 
 The film profile, defined by
the isoline at density $\rho=\rho_{bulk}/3$, splits in two regions. In
the first one, reaching up to $y\approx7 \Delta z$, the front is
fitted by the expected bulk circle profile.  In the second one,
extending from $y\approx7 \Delta z$ to the wall, the film profile is
fitted by the analytical  expression $\sim Ca^{-1}a^2/z$, where $a$ is
a characteristic molecular size, which in our case is set equal to
$a=1$.  The analytical profile was proposed for the case of ``maximal
film'' (perfect wetting)~\cite{DeG_84}, under the lubrication
approximation, with Van der Waals interactions between fluid and walls.
Quantitative agreement is again found between LB, MD and the theoretical treatment.

 In order to further analyse the previous results, we  study the asymptotic case, i.e. a situation with a very small fluid
velocity (quasi-static), obtained in a longer channel ($L=nz=1500  \Delta z$) with the same geometry
depicted in fig. \ref{Fig:1}. 
To be noted that such domain is too large for an atomistic method and,
therefore, we have simulated it with LB only.
In such a case, the departure between the bulk profile and the
precursor shape is even more pronounced, due to the longer
elongation of the film close to the wall. In the inset of the same
figure, we also show the macroscopic angle formed by the interface as a
function of the distance from the wall.  The bulk interface is fitted
by a circle extrapolating to the wall with a contact angle
$\theta=0$.
It is remarkable that the presence of the precursor film allows the apparent contact angle
to vanish less sharply than in the ideal circular case, as visible in the inset of fig. \ref{Fig:4}b.
The deviation from the circular fit can not be  due to a simple dynamical distortion 
of the  apparent contact angle, since this would lead to an increase of the macroscopic angle.
Therefore, this deviation can be interpreted as a signature of the onset of  precursor physics.

Summarising, we have reported quantitative evidence of the
formation and the dynamics of  precursor films in capillary filling
with highly wettable boundaries. The precursor shape shows 
persistent  deviation from an ideal circular meniscus, due to
the nanoscopic distortion induced by the interactions with the
walls. This is likely connected to the disjoining pressure induced
by van der Waals interactions between fluid and solid~\cite{DeG_book}. The precursor
follows the same Lucas-Washburn functional law as the bulk meniscus,
although  with a
different prefactor, see eq. (\ref{eq:prec}).
This prefactor has been found to be the same for both geometries investigated,
while its universality with respect to other physical parameters remains to be explored.
Our findings are supported by direct comparison between LB,  MD simulations and
theoretical predictions.
On the practical side, the hydrokinetic method  
presented in this work offers the opportunity 
to perform very efficient numerical simulations 
of precursor formation and propagation in 
complex geometries of direct interest for the optimal design of micro- and nanofluidic devices.
Furthermore, a very intringuing result is that the precursor film near-to-the-wall 
is found to cause a distorsion of 
the apparent contact angle, which suggests
the possibility to design devices with increased efficiency in presence of roughness.


\end{document}